\documentclass[12pt,a4paper]{article}
\usepackage{hyperref,graphicx,amsmath,amssymb,lineno,array,color}
\usepackage{cite}
\usepackage{geometry}
\usepackage[footnotesize,bf]{caption}
\date{}

\begin{document}

\title{Reciprocating thermochemical mediator of pre-biotic polymer decomposition on mineral surfaces}

\author{ Rowena Ball\thanks{Corresponding author. Email Rowena.Ball@anu.edu.au, Orcid ID 0000-0002-3551-3012}\\
Mathematical Sciences Institute \\  Australian National University,  Canberra 2602, Australia. \\
\and
John Brindley\thanks{Email J.Brindley@leeds.ac.uk.}\\
School of Mathematics, University of Leeds, Leeds LS2 9JT, U. K. }
\maketitle

\begin{abstract} \noindent \raggedright
A continuing frustration for origin of life scientists is that abiotic and, by extension, pre-biotic attempts to develop self-sustaining, evolving molecular systems tend to produce more dead-end substances than macromolecular products with the necessary potential for biostructure and function --- the so-called `tar problem'. Nevertheless primordial life somehow emerged despite that presumed handicap.   A~resolution of this problem is important in emergence-of-life science because it would provide valuable guidance in choosing subsequent paths of investigation, such as identifying pre-biotic patterns on Mars. 
To study the problem we set up a simple non-equilibrium flow dynamical model for the coupled temperature and mass dynamics of the decomposition of a polymeric carbohydrate adsorbed on a mineral surface, with incident stochastic thermal fluctuations. 
Results  show that the model system behaves as a reciprocating thermochemical oscillator. The output fluctuation distribution is bimodal, with a right-weighted component that guarantees a bias towards detachment and desorption of monomeric species such as ribose, even while tar is formed concomitantly.
This fluctuating thermochemical reciprocator may ensure that non-performing polymers can be fractionated into a refractory carbon reservoir and active monomers which may be reincorporated into better-performing polymers with less vulnerability towards adsorptive tarring.  

\bigskip \noindent 
Keywords: Emergence of life, Tar problem, Pre-biotic, Thermochemical oscillator, Surface molecular processes, Stochasticity

\end{abstract}

\setlength{\parskip}{6pt}

\newpage

\section{Introduction} 
It is well-accepted that the organic geochemical processes from which living structures emerged on the primordial Earth must have operated episodically, at least, if not quasi-continuously, in a far-from-equilibrium milieu. In thermodynamic terms an open flow system is required \cite{Cupic:2019}. 
 Model systems for studying pre-biotic molecular sub-processes usually are maintained, or assumed  to operate,  at steady state, where the rates of inflow (of raw materials or molecular building blocks) and outflow (of more complex products and unreacted species) balance the chemical reaction rates for all time. However, we have established previously that non-negligible mass and enthalpy \textit{dynamics} are necessary to mediate  net polymerization and molecular evolution and amplification in solution; a far-from-equilibrium steady state process ultimately will be unsuccessful \cite{Ball:2020,Ball:2021}. Effectively this means that   internal oscillatory dynamics and  environmental stochasticity should be sustainable in the medium. Here we extend those results to include heterogeneous surface processes, for which environmental conditions on the primordial Earth may have been supportive.  
 
 Researchers of pre-biotica, understandably, have been more interested  in how  synthesis, polymerization and proliferation of biomolecules may have occurred rather than their decomposition, and important advances have been made on syntheses of  ribose, nucleic and amino acids, and fatty acid esters  without enzyme catalysis \cite{Yadav:2020, McCollom:2013b}. But the inconvenient fact remains that attempts to develop abiotic and, by extension, pre-biotic  self-sustaining, evolving molecular systems tend to produce more dead-end substances than macromolecular products with potential for biostructure and function. This is the so-called ‘tar problem’. 
 
 Notwithstanding this inevitable nuisance, the notion that oligomerization of monomers and molecular self-organization 
ought to be steadily progressive  is  beguiling and persistent \cite{Harrison:2023}. Consequently, various ways by which precious, fragile biomolecules could be concentrated and protected from decomposition  have been investigated. Some non-mineral based protection mechanisms proposed include sequestration from the hydrolytic milieu by coacervation \cite{Ghosh:2021}, simple amphiphile encapsulation  \cite{Cohn:2004,Ruiz:2014}, finding refuge in the interstices of chondritic insoluble organic matter \cite{Ischia:2021}, and associating with tars made from the biomolecules themselves \cite{Saha:2022}.
However, mineral--fluid interfaces have been widely accepted as the setting for this supposed monotonic simple-to-complex progression, being capable of providing concentration, protection, and catalytic capabilities that enabled molecular pre-biotic systems to grow, persist, self-organize and diversify \cite{Bernal:1949,Hazen:2006,Wachterhauser:2006,Martin:2008,Cleaves:2012,Hazen:2017,Damer:2020,Westall:2023}. 

Taking stochastic periodicity and a mineral-based interfacial flow setting as necessary conditions therefore, we are led  to consider whether stochastic surface thermochemical oscillations may provide the dynamics to maintain a pre-biotic molecular flow system in non-steady-state. Thus, in this study, we interrogate  a dynamical model for a simple thermochemical process on mineral surfaces, involving localised thermal feedback and stochasticity, which effects the conversion of adsorbed organic oligomers alternately to surface-bound carbonized products and to free organic monomers. The process performs as a reciprocating oscillator\footnote{We adopt the descriptor `reciprocating' and the term `reciprocator' by analogy with a mechanical reciprocating oscillator where some of the power output of one stroke is fed into the alternate stroke.}  that allows organic species to be recycled, but also sequesters refractory carbon in a long-lived reservoir. 

\newpage
Thermochemical oscillations on mineral and metal surfaces have been widely observed and intensively studied from the 1970s onwards \cite{Sheintuch:1977}, mostly in the context of  heterogeneous catalytic processes for various industrial and technological applications. In a 1993 review  Schuth et al.  \cite{Schuth:1993} commented that
\begin{center}
\begin{minipage}[c]{0.9\textwidth}
`Oscillations may be lurking in every heterogeneous catalytic system, and one might speculate that every such reaction might show oscillations under the appropriate conditions.' 
\end{minipage}
\end{center}

The growing corpus of research prior and since, on an ever-larger variety of surface-reactive systems, endorses this statement \cite{Wicke:1986,Vlachos:1992,Zhang:2005,Slinko:2001,Cordonier:1989,Slinko:2020,Kipnis:2021,Raab:2023,Peskov:2023}. Observed oscillations of surface-reactant concentration may be nominally isothermal, occurring due to chemical feedback giving rise to power-law rate chemistry that is effectively third- or higher-order,  and/or non-isothermal, where wall thermal conductance is non-negligible and thermal feedback is exponential due to the temperature dependence of rate constants. The surface reactions studied almost always  are degradative or oxidative, and the design of such systems for chemical engineering applications usually aims to achieve efficient conversion of a hydrocarbon adsorbate  with minimization of the inevitably accompanying surface carbonization and consequent site blocking or deactivation (catalyst poisoning)  
\cite{Bychov:2018,Peskov:2023}. 

Some fraction of the native surface reactions in a pre- or proto-biotic setting also must have been degradative, producing surface carbonization or insoluble organic matter (IOM, which here we use synonymously with tars, chars, and asphalt, or more bluntly, dead-end products) \cite{Ischia:2021,Sponer:2021}. Molecular degradation is an essential feature of fully-fledged biology --- for that, of course, is the end-game of metabolism --- and it is well-understood that the cycle of life includes the decomposition of polymeric species and re-use of monomeric decomposition products. By continuity argument, it is reasonable to assume that chemical decomposition and recycling also were unavoidable in  the pre-biotic world --- and, indeed, essential to its  functioning. 
 As a body of knowledge, though, emergence-of-life science tends to  downplay or sidestep the realities that biomolecules are in continuous cycles of synthesis and degradation, and that to achieve net growth we simply need synthesis to occur slightly faster than degradation in an open system far from equilibrium. The emergence of a `kinetic trap' when the rate of synthesis exceeds that of the back reaction was recognised by Ross and Deamer \cite{Ross:2016} and Damer \cite{Damer:2020}. 

It is perhaps no wonder that many researchers strive for purely synthetic pre-biotic scenarios, but, as intimated above, in heterogeneous abiotic/pre-biotic setups, the catch is that molecular decomposition also occurs and is \textit{always} accompanied by  tar or IOM formation. 
The `asphalt problem' or `tar paradox' \cite{Benner:2014,Benner:2023} of the RNA world --- or of any hypothesized and experimentally supported pre-biotic chemical `world' --- has been treated as a major annoyance, particularly by experimenters: reactive mixtures of biomolecular precursors, such as (poly)nucleotides tend to devolve into complex, highly conjugated (and therefore dark in colour), and biologically useless  tars. 

   \begin{figure}[t]
 \centerline{
 \includegraphics[scale=1]{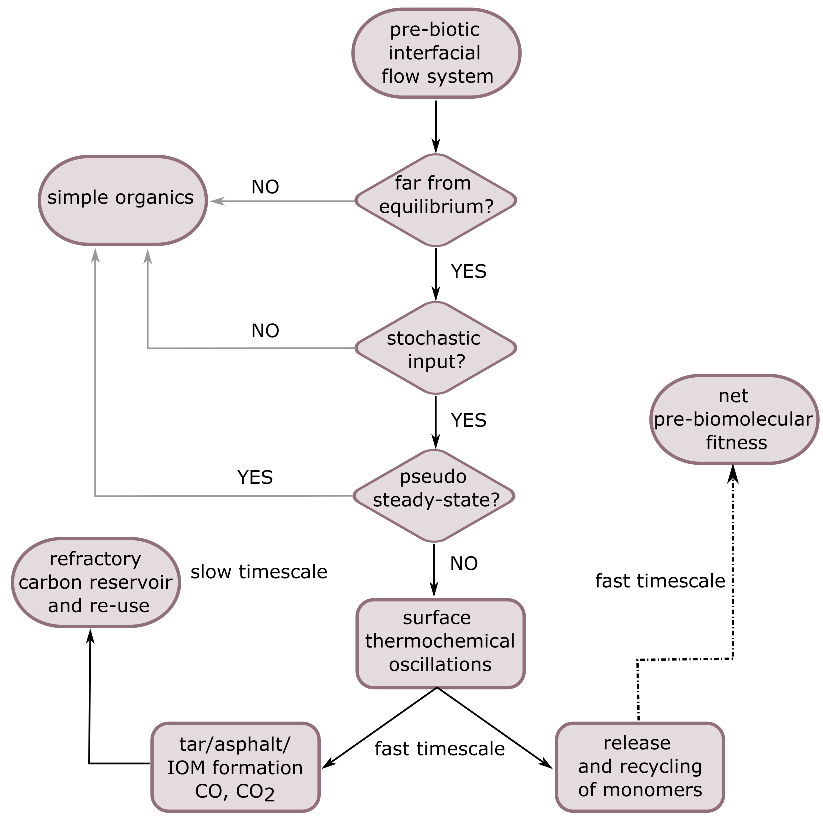}
 }
 \caption{\label{figure1} Flowchart summarizing the key decision criteria and processes discussed in the text that enable the thermochemical dynamics that lead to the reciprocating formation of tars and release of monomers. The flowchart is discussed further in section 4.}
 \end{figure}

The `tar paradox' was confronted by Benner et al. \cite{Benner:2012,Neveu:2013} who proposed and studied borate as an effective  mediator and complexing agent by which pre-enzymatic carbohydrate chemistry may have dodged --- or at least managed or mitigated --- the asphalt or tar problem. Mamajanov \cite{Mamajanov:2019}  found that onset of tarring of branched polyesters can be delayed by wet-dry cycling, while Col\'on-Santos et al. \cite{Colomer:2020} cycled formose reactants and formamide over minerals to help alleviate (somewhat)  the formation of intractable messes.

But is asphaltization really a problem at all? Perhaps it is an essential element of the organic geochemistry that gave rise to self-perpetuating molecular systems. In the `nuclear geyser' model for emergent biochemistry proposed by Ebisuzaki and Maruyama  \cite{Ebisuzaki:2017} tar is not a dead-end material but has an important role as a reservoir of biomolecular substrates and intermediates, which are released by radiolytic processes. 

In fully-fledged  biology, we simply do not observe asphaltization: enzymes seem to have taken care of that problem admirably. But we are interested in pre-biotic tars because, firstly, their evidence could serve as a  sign of past incipient life that failed and, secondly,  they may provide reservoirs of reduced carbon that could be transported by geophysical processes or by meteorites and possibly mined by ensuing generations of pre-enzymatic chemistry.   

In this study our results suggest  a possible role for stochastic thermal surface oscillations in facilitating molecular complexity (this is discussed in Section  4), but also belie the  conceit  of the emergence of life as a steadily progressive continuum,  as they suggest important roles played by degradation and by carbonized products  and IOM and support the value (to researchers) of false starts and dead ends \cite{Smith:2021}. 
The following sections develop and present an in-principle demonstration of how an inherent surface thermochemical reciprocator may have functioned advantageously towards the emergence of life.  We show that the enthalpy stored and released by reciprocating surface thermal oscillations can power the release of monomers from non-performing polymers for subsequent recycling into (possibly) better polymers.   Refractory carbon is an inevitable, but not necessarily a nuisance or detrimental, byproduct. 

The above introductory and motivational paragraphs are summarized in the flowchart diagram of figure \ref{figure1}, which will be addressed further in section 4.

\section{Simple model for thermochemical dynamics on mineral surfaces}

We consider a  surface-reactive system of a layered or porous mineral where the interfacial flow may carry free organic species and fine particles --- probably clays \cite{Hazen:2017} --- with adsorbed oligomers. In the large, such a system is complex in solid structure and fluid advections~\cite{Lester:2013}, even without chemical reaction and thermal dynamics, hence we define a simple subsystem using the CSTR (continuous-flow stirred tank reactor) paradigm. 
Physically, the CSTR may be thought of as a single pore to  which the inflow provides a continuous excess supply  of `clean' surface (i.e., uncontaminated by carbonized or charred material) with oligomers adsorbed on clay mineral particles in suspension.

As the adsorbed oligomer we single out the complex, branched, and  stereochemically impure carbohydrates produced by autocatalytic condensations of formaldehyde, in a complicated process usually just called `the formose reaction'   
(where it is understood that formose is not a substance but a descriptor). 
This reaction has been studied widely in pre-biotic scenarios, because  one of the many products, ribose, is of indisputable importance to life, being a component of the nucleotides \cite{Yadav:2020}, a mediator of nucleotide synthesis  \cite{Tran:2023}, and providing starting materials for synthesis of other biomolecules such as the aromatic amino acids \cite{Camprubi:2022}.  The formose reaction also is feasible at moderate temperatures under product cycling (recursion) \cite{Colon:2019}  and in wet-dry cycling conditions \cite{Damer:2020}. We cannot say for certain that ribose was involved in the emergence of life, although evidence of extraterrestrial ribose has been reported in carbonaceous chondrites \cite{Furukawa:2019}. Mechanochemical formation of ribose (and other sugars), catalyzed by minerals, has been demonstrated and proposed to have occurred on the early 
Earth \cite{Haas:2020}, and a variety of sugars  has been produced from the formose reaction in the presence of olivine silicates as catalysts \cite{Vinogradoff:2024}.

However, the formose reaction is infamously non-selective: the ribose monomer typically is only a minor monomeric sugar product, branched polymeric carbohydrates also are formed, the retroaldol reaction can occur \cite{Nogal:2023}, and in the presence of various minerals the formation of tars, chars and IOM is observed if allowed to proceed in an autocatalytic cascade \cite{Orgel:2004,Schwartz:2007}. In the presence of amine species, the formose reaction products also undergo Maillard reactions, also leading to tarry solids.

\begin{figure}[t]
 \centerline{
 \includegraphics[scale=1]{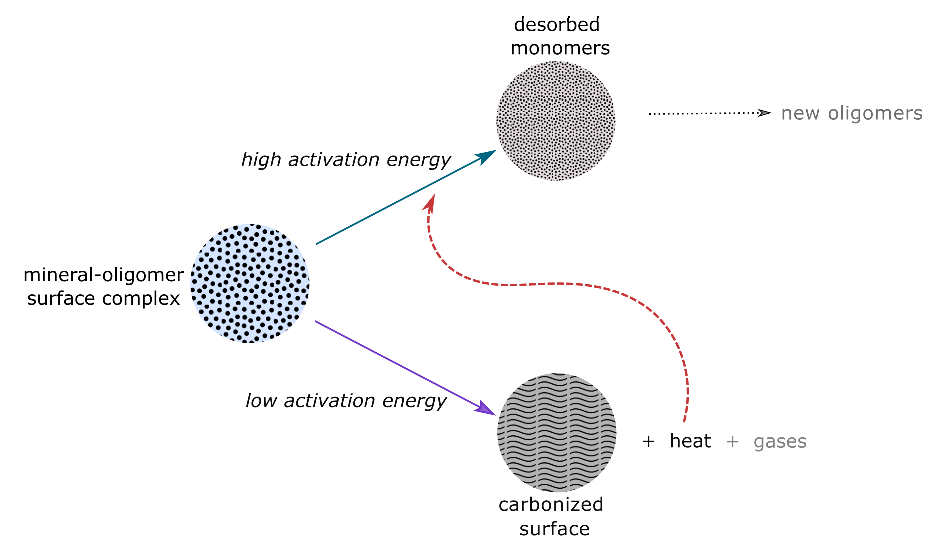}
 }
 \caption{\label{figure2}Operational schematic of the surface thermochemical reciprocator: The heat produced by the low activation energy, hence predominantly low temperature, carbonization/tarring process feeds back to activate the high-temperature detachment and desorption of monomers, which locally cools the system thus switching it back to the alternate process. Carbonization also produces gases such as CO, CO$_2$, and water vapour. Desorbed monomers may form new oligomers, possibly on the same type of surface.}
 \end{figure}

Within the CSTR paradigm, where wall thermal conductance is restricted locally by the solid mineral structure, it is reasonable to treat the thermal decomposition of adsorbed carbohydrate oligomers as being similar to the well-characterized thermal decomposition of cellulose  \cite{Ball:2004,Sullivan:2012}. As illustrated in  figure \ref{figure2}, 
two thermally coupled processes may occur: exothermic reactions in the solid phase,  
and an endothermic reaction, which involves detachment and desorption of a monomeric species. The thermal coupling of exothermic and endothermic reactions occurs through the Arrhenius form of the reaction rates, thus the temperature is dynamically variable.
(This is a special case of an Endex system \cite{Ball:1999}.) The process may be conceptualised as a reciprocally acting couple, for which alternating high and low temperature states are intrinsic. From the minimalist scheme in figure \ref{figure2} the dynamics of its operation may be deduced  as follows:

\begin{itemize}
\item Over a relatively low temperature range, slow dehydration and carbonization (`tarring' or `charring') of the surface organic complex takes place. These are low activation energy, exothermic  reactions. 
\item The substantial heat  produced by the slow carbonization reactions increases the local temperature. The resulting  temperature spike transiently favours the alternate reaction, a high activation energy  detachment and desorption of a monomeric species into solution.
\item Desorption, however, is endothermic and as the site cools  the low activation energy  process takes over again, perpetuating the cycle. 
\end{itemize}
There is a relentless inevitability about the action of this built-in  reciprocator, given a continuous supply of oligomer-saturated sorption sites, or of insoluble polymer. 
Although the thermal oscillations that manifest may be damped out by other linear (pore wall) or nonlinear (reactive) heat transfers, where they are substantiated, even intermittently, in the pre-biotic setting there is potential for the released monomeric species to be incorporated subsequently  into new, possibly fitter, polymers or mediate the formation of nucleotides and some amino acids. 

The dynamics of this heterogeneous thermochemical oscillator were first elucidated for the thermal decomposition of cellulosic species \cite{Ball:2004}, where the `surface' is the cellulose chain itself (often complexed with metal ions), the analagous adsorbate is a positively charged thermolysed cellulose end, and water or acid may promote the charring stroke. The chemistry at the heart of this process was elucidated in \cite{Sullivan:2012}.

In view of these studies, and the experimental literature on thermochemical oscillations cited in section 1, the presumption here  that similar chemistry and dynamics may operate on mineral surfaces coated with adsorbed amorphous carbohydrate species is well-founded, if not experimentally observed to date. Qualitatively comparable models, informed by experimental data, developed to describe oxidation  
of adsorbed gas species on solid catalysts were reviewed in \cite{Schuth:1993}.  Peskov and Slinko \cite{Peskov:2023} have explicitly described  and modelled a similar reciprocating thermochemical cycle for methane oxidation on nickel, which provides valuable substantiation for our model and its assumptions. A salient point of difference is that the surface carbonized material is oxidized at high temperature (c. 1330\,K) in their model, whereas over the moderate temperature range of our model that rate is reasonably assumed to be zero, and the material persists as tar --- that, after all, is the essence of the `tar problem'.
  
 In the CSTR paradigm the coupled mass/heat evolution implied by the schematic in figure~\ref{figure2} are minimally captured by the following dynamical system. 
 \begin{align}
    \bar{C}\frac{\text du}{\text d\tau}& = e^{-1/u} + \alpha \nu e^{-\mu/u} + \left(\varphi \bar{C}+\ell\right)\left(\tilde{u}_{\text a} - u\right) \label{e3}\\
    \frac{\text dw}{\text d\tau} &=  -e^{-1/u} + \gamma + \varphi\left(w_\text{f} -w\right) \label{e1}\\
  \frac{\text dx}{\text d\tau} &=   \nu e^{-\mu/u}  - \varphi x \label{e2}. 
 \end{align} 
For clarity of presentation and convenience of computation, equations \eqref{e3}--\eqref{e2} are cast in terms of dimensionless quantities, which are defined in Appendix table \ref{table1}. The dependent variables are the temperature  $u$, the amount of desorbed monomer $x$, and the amount of dehydration water $w$ produced and consumed by tarring reactions, which serves as a convenient proxy for the solid carbonized residue. The right-hand side terms in the dynamical enthalpy balance equation \eqref{e3} represent, respectively, heat production by exothermic reaction, the heat requirement of the endothermic reaction (note that the parameter $\alpha$ is negative), and heat losses to the environment at temperature $\tilde{u}_\text a$ by flow and by thermal conductance through the solid medium or pore walls.

It is particularly important to note the following distinction of this model: The temperature $u$ is a \textit{dynamical} variable and is \textit{self-consistently} coupled to the reactant concentrations through the exponential temperature dependence (Arrhenius) of the reaction rates. The thermally self-consistent equations \eqref{e3}--\eqref{e2} simply encode the dynamics schematized in figure \ref{figure2}, where the charring reaction heat kicks the high-temperature desorption reaction, which then cools and switches the system back to the low-temperature charring. No external temperature cycling or forcing by an experimenter is necessary. This self-sufficiency also means that possible geophysical sites for which the model is valid are not limited to those for which an external driver is required to maintain thermal cycling.  

\newpage

Equations \eqref{e3}--\eqref{e2} are based on those derived in \cite{Ball:2004} for cellulose thermal decomposition, where the released monomer was levoglucosan and the solid residue was char. Here we also have incorporated the effects of incident thermal stochasticity by setting the ambient (environmental) temperature $\tilde{u}_{\text a} $ as a normally distributed random quantity, 
\begin{equation} \tilde{u}_{\text a} =  \bar{u}_{\text a} + \delta u_{\text a}, \label{e4}\end{equation}where where $ \bar{u}_\text{a}$ is the mean and $\delta u_\text{a}$ is the fluctuation with variance $\sigma$. This model for a randomly fluctuating environmental temperature was used by Ball and Brindley in \cite{Ball:2017,Ball:2019,Ball:2020}, where the non-intuitive --- but crucially important for the emergence of life --- result was obtained that a randomly perturbed (or stochastic) input temperature on a dynamical thermochemical system produces \textit{non-randomly} distributed output fluctuations.  

Equations \eqref{e3}--\eqref{e4} were solved numerically  using a stiff integrator. First, the fluctuation $\delta u_\text{a}$ was turned off and the parameter space was searched and mapped for oscillatory regimes using numerical eigenvalue (linear stability) analysis. Using the sign of the eigenvalues as a guide, we could choose initial conditions of temperature $u$ and reactant amounts $x$ and $w$ that produce trajectories that settle on stable limit cycles or on stable steady states --- after which the initial conditions are `forgotten'. Then long time series were taken with $\delta u_\text{a}$ turned on, to obtain robust statistics. (For a primer on eigenvalue analysis see \cite{Jordan:2007}.)

For the reactive system coded in equations \eqref{e3}--\eqref{e4} the assumptions made are severe but sound, and can be valid approximations. We make the `infinite pool' assumption for the reservoir of oligomer-coated mineral substrate ---  for example, a large excess of surface-bound oligomer on clay particles in suspension in the inflow to the CSTR ---  as did W{\"a}chtersh\"auser \cite{Wachterhauser:2006} with respect to the postulated pre-biotic iron-sulfur world. The desorbed product is taken as a single monomeric moiety, and the usual CSTR assumptions apply. These conditions are achievable by experimental design, and in the `pre-biotic wild' we are free to choose and isolate a small volume where the approximations made for the purposes of a defined study are valid (note that Hazen \cite{Hazen:2017} makes a plausible estimate of the range of possible `experiments' in pore-sized CSTRs in the ranges of time and space available in the prebiotic world). Similar assumptions were made by Ord et al.  \cite{Ord:2012} for geochemical modelling of reactive hydrothermal mineralized systems. 
We note, too, that the use of experimental flow reactors with attached diagnostic instruments  to investigate pre-biotic mineral-organic chemistry  has been developed by Kawamura \cite{Kawamura:2017}, as an analogue of natural submarine hydrothermal vent environments. However those experiments were thermostatted to a target temperature so thermochemical dynamics was not studied in real time, as we have done here in simulation. 

\newpage
\section{Results}
 
\begin{figure}[t]
 \centerline{
 \includegraphics[scale=1]{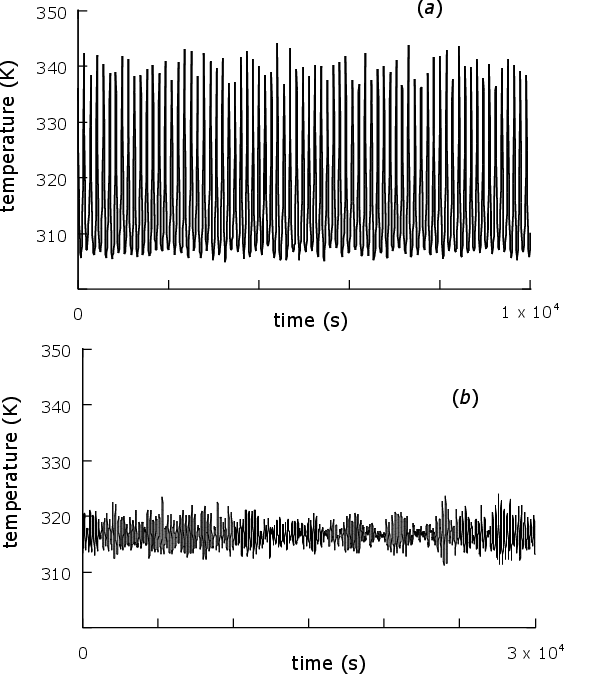}
 }
 \caption{\label{figure3} Stochastic time series windows for (a) an oscillatory state, and (b) a pseudo steady state.  Data were computed from equations \eqref{e3}--\eqref{e4} with $\mu=E_2/E_1=1.7$ and  fluctuation variance $\sigma=1.7\times 10^{-3}$. See \href{https://doi.org/10.5061/dryad.kh18932hr}{https://doi.org/10.5061/dryad.kh18932hr} for code and other parameters.}
 \end{figure}

\begin{figure}[t]
 \centerline{
 \includegraphics[scale=1]{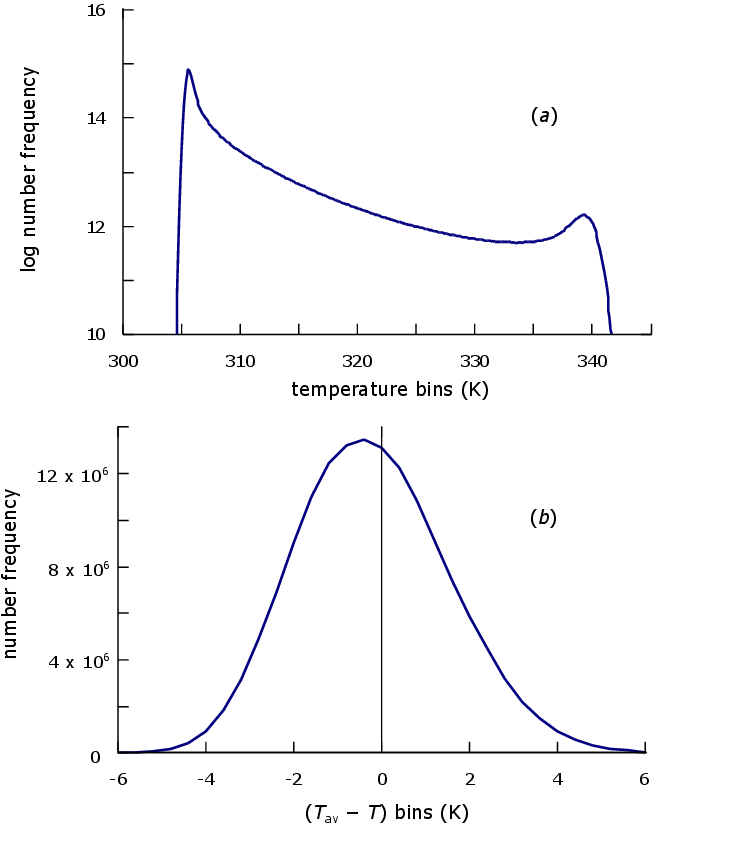}
 }
 \caption{\label{figure4} For the histograms above, the data for integration time of $3\times 10^6$\,s computed with a time step of 0.02\,s were  sorted into 100 temperature bins, the smoothed envelope of which is shown. (a) Distribution of the stochastic oscillating temperature in figure \ref{figure3} (a) is distinctly bimodal, as accentuated by use of a log scale for the number frequency. (b) Distribution of the stochastic pseudo steady state temperature  in figure~\ref{figure3} (b) binned as deviations from the mean, $T_\text{av}-T$, is distinctly  left-weighted. }
 \end{figure}

The essential computation results are presented in figure \ref{figure3} and figure \ref{figure4}, where the output data have been re-scaled to the corresponding dimensional quantities. The kinetic parameter ratios $\mu$ and $\nu$ were selected to fall within ranges suggested by the values given in table \ref{table2} in the Appendix. Although these ratios do not correspond to measured kinetics for the surface carbohydrate reactions considered here (since such data are not available), the perturbed oscillatory behaviour is robust over a wide range. 

In figure \ref{figure3} stochastic time series are rendered for (a) an oscillatory state and (b) a pseudo steady state. It can be seen that the temperature range, in this case, is mild and conducive to aqueous organic reactions of biochemical relevance, and that the period of the oscillation in (a) is of order 2--3 minutes. To obtain the pseudo steady state in (b), the oscillatory dynamics was suppressed by increasing the specific heat parameter $\bar{C}$, which multiplies the dynamics (the left hand side) in equation \ref{e3}, until the eigenvalues became negative. Although an artifice in physical terms, for comparative purposes this ensures that the \textit{average} temperatures in (a) and (b) are equal, which allows us to obtain from the computed data the  result given in table \ref{table0}. 

\newpage 
Experimentally measured temperature amplitudes in the chemical engineering literature for thermochemical oscillations range from 10--500\,K (see, for example, refs  \cite{Stavarek:2015,Sheintuch:1977,Schuth:1993}), so we see no technical impediment to the actuation of well-designed experiments aimed at observing the behaviour shown in  figure \ref{figure3} (a).

The incident, or input, thermal stochasticity --- i.e., normally distributed $\delta u_\text{a}$ in equation \eqref{e4} --- models the inevitable perturbations of temperature, density, or chemical potential in any natural environment. What do the distributions of the \textit{output} thermal fluctuations  in figure   \ref{figure3} look like? 
Histograms of the data from extended time series are presented in figure \ref{figure4}, where in each case 100 bins were used and the envelope is outlined in blue. 

\begin{table}[t]\caption{\label{table0} The long-term time average amount of released monomer $\bar{x}$ for a stochastic dynamical (oscillatory) state is some 12\% higher than that for the equivalent steady state, and the average amount of the tar-proxy $\bar{w} $ is correspondingly lower.}
\centerline{
 
 \begin{tabular}{lccc}
 \hline  
& $\sigma$ & $\bar{x}$ & $\bar{w}$ \\
\hline
steady state &$1.7\times 10^{-3}$& 0.203 & 0.800\\
  		 & 0 & 0.202 & 0.800\\
dynamical state &$1.7\times 10^{-3}$& 0.228 & 0.772\\
			& 0 & 0.224 & 0.776 \\
\hline
\end{tabular}
}
\end{table}

In figure \ref{figure4} (a) the raw temperature data from which figure \ref{figure3} (a) was rendered were sorted into the bins so that  log number frequency could be used for display, to accentuate the distinctly bimodal distribution. This bimodal distribution was found previously \cite{Ball:2020} for a solution-phase thermochemical oscillator; here we have confirmed that it also occurs for surface oscillations. The high temperature component is the distribution of the oscillation maxima. The presence of this component creates an inbuilt bias, ensuring that the high activation energy reaction, detachment and desorption of a monomer, is strongly sustained, even while tar/IOM formation inevitably occurs at low temperatures. 

The histogram in figure \ref{figure4} (b) of the pseudo steady state temperatures in figure \ref{figure3} (b) is rendered in terms of deviations from the average temperature. In contrast to (a) the distribution is single-peaked and left-weighted, thus the frequency of high temperature fluctuations is relatively poor. This distribution tells us that a stochastic steady state strongly favours low temperature, or low activation energy, reactions. These are the tar/IOM formation reactions.  
  
\section{Discussion}
For the purposes of this work we take the existence and supply of carbohydrates as given\,---\,in other words, the formose or other reactions provide a continuous source. The circulation of these and other organic species in an interfacial pre-biotic setting must have involved both the  formation and decomposition of oligomers, at least episodically or intermittently if not continuously. Most primary research in this area has been concerned with pre-biotic (non-enzymatic) polymerization, with scant recognition of the essential role of decomposition. Interesting exceptions include studies of the decomposition of norvaline   \cite{McCollom:2013} and the hypothesis proposed in \cite{Yakhnin:2013} that pre-biotic cycling between monomers and polymers could concentrate macromolecules at mineral surfaces.

As with synthesis, pre-biotic decomposition of polymers must be achieved non-enzym\-{atically} and at moderate temperatures. Charring and browning of organic substances usually is associated with rapid conversion at high temperatures, or pyrolysis, but these processes do take place at moderate temperatures, albeit  more slowly. Experimental chemists routinely observe the (usually unwelcome) tarry products of Maillard reactions in their flasks, and librarians and archivists are concerned with the yellowing and browning of historic papers, palm leaves, parchments and vellums. In the geophysical/geochemical environment it has been confirmed  that abiotic Maillard reactions  at moderate temperatures, catalysed by iron and manganese minerals, can geosequester enormous amounts of organic carbon \cite{VanBoekel:2001,Moore:2023}.  

Here we have shown that in a pre-biotic mineral interfacial setting, adsorbed oligomeric sugars may be partitioned into a refractory carbon reservoir (tar, char, IOM) and a supply of monomers desorbed continuously into the inner aqueous boundary layer or into solution. The monomeric species  released may be adsorbed onto  conducive neighbouring mineral sites \cite{Bhatt:2022}, 
or recycled into new populations of perhaps fitter polymers, such as polyribonucleotides. Although we have not complicated and extended our simple model in this work to include recycling, W{\"a}chtersh\"auser \cite{Wachterhauser:2006} describes this process vividly: monomeric species that are detached from the surface may diffuse in the inner sphere, or boundary layer, of the water, and may be re-captured and salvaged downstream by species on the surface, or they may be lost to the system in bulk solution. Those species that are lost in solution are less likely to be useful, thus a certain amount of self-selection occurs. (We note, however, that tarring evidently does not occur in W{\"a}chtersh\"auser's pre-biotic world.)  
In this context `fitter' polymers are those which are more functional and less likely to be fated to join the refractory carbon reservoir. 

The thermochemical reciprocator process thus fulfils a very basic sorting and winnowing function, effectively a primitive form of selection. If we envisage a flow through successive pores of a porous matrix (which has not been modelled here), each fluid particle, together with its suspended polymeric cargo, may provide a succession of `rebirths' of monomers for recycling.
New ribonucleotide oligomers formed using released monomeric ribose  may add to existing oligomers or catenate into new polymers downstream, given a power supply --- possibly via redox reactions of hydrogen peroxide, which is also produced on mineral surfaces  \cite{Borda:2001,He:2021} and, as a thermochemical oscillator in its own right, has been shown in simulation to power reactions of pre-biotic relevance \cite{Ball:2014,Ball:2015,Ball:2017,Ball:2020}.  Some of the newly-formed polymers may have grown faster and longer, aggregate better, and form inter- and intra-molecular structures, thus avoiding surface carbonization. Other, poorly-performing polymers may be `sent' for surface decomposition and recycling/carbon sequestration. 

Returning to figure \ref{figure1}, these processes are schematized in the flowchart. The first three decisions tell us whether a pre- or protobiotic interfacial flow system is capable of dynamic behaviour and can exhibit the reciprocating thermochemistry described in figure \ref{figure2}.  A system near equilibrium, without stochastic input, or operating at steady state must revert to simple, monomeric organic molecular species. Otherwise, surface-mediated thermochemical oscillations are intrinsic and inevitable. On the fast timescale of the oscillations tar formation and monomer release and desorption occur reciprocally. Downstream, on the fast timescale, the monomers may be recycled, and, on the slow timescale, the reservoir of reduced carbon formed may be mined --- these processes are not modelled in this work.   

Isothermal chemical oscillations of organic reaction networks to demonstrate  pre-biotic metabolisms have been modelled by Papineau et al. \cite{Papineau:2023}, and were studied using CSTR experiments  by Semenov et al. \cite{Semenov:2016}. Their experiments were carried out in solution under thermostatted conditions and the  feedback was effected by chemical autocatalysis. Autocatalysis is also likely to have been effected by thermal feedback, as we have shown in this work, schematized in figure \ref{figure2}. In general both thermal and chemical feedbacks may be involved in creating and driving a complex non-enzymatic molecular system. 

However, we note that isothermality (i.e., thermostatting) is a condition imposed externally by an experimenter. In our self-consistent thermochemical system, equations \eqref{e3}--\eqref{e2}, the temperature evolves naturally as a dynamical variable. Moreover, even if the system were to operate in approximate isothermal mode --- perhaps episodically as the flow accesses a region of anomalously high wall thermal conductance (the parameter $\ell$, equation \eqref{e3}) in the porous matrix --- the incident fluctuation itself, equation \eqref{e4}, would \textit{not} be isothermal, hence thermal dynamics would still govern the system.  

The purely thermochemical process we have described shows how  tars are inevitably produced from some fraction of the adsorbed oligomers ({figures \ref{figure1} and \ref{figure2}). Tars are nuisance `dead end' products according to Benner \cite{Benner:2014} but their potential as prebiotic sources of useful organic species has been explored by several groups \cite{Schwartz:2007,Lavado:2016,Juntunen:2018,Mamajanov:2019,Ischia:2021,Sponer:2021,Kopacz:2023}, and it has been suggested that non-biomolecular organic condensates may have been important for the origin of life \cite{Jia:2023}.  As solids complexed with minerals, refractory carbon substances have been carried around the solar system in meteorites, possibly serving as tarpits (as it were) of useful material at the dawn of life. Refractory carbon and IOM also oxidizes, and may undergo non-aerobic oxidation, with minerals such as Fe(III) oxide serving as the electron acceptor, or photochemical decomposition \cite{Kopacz:2023}.  In a pre-biotic scenario the CO$_2$ liberated may participate in synthetic reductions to methane, formaldehyde, and glycoaldehyde. 

Although we have exemplified the model presented here with an adsorbed carbohydrate polymer releasing monomeric ribose under thermal decomposition, as having relevance to the pre-biotic growth and persistence of oligomeric ribonucleotides, tar products result from the thermal decomposition of many types of organic polymers. The Maillard reaction involving polymers with amide or amine groups is a  well-known tar producer, and it is a fair supposition that pre-biotic Maillard reactions would have occurred with surface-adsorbed peptides.  The manifestation of reciprocating oscillations (figure \ref{figure3} (a)) depends on the thermokinetic parameters and the heat loss rate. Exemplary thermokinetic parameters for a number of pre-biotically relevant species are collected in the Appendix table~\ref{table2}. The heat loss rate coefficient, $\ell$ in Appendix table \ref{table1}, is a parameter that may shore up thermochemical oscillations over a certain range in pre-biotic nature or be adjusted optimally by the experimenter. 
 
Our results throw a positive light on Benner's `tar problem' --- in a continuous flow, far-from-equilibrium, non-steady-state system, the bimodal distribution of output thermal fluctuations shown in Figure \ref{figure4} (a) will ensure production of useful monomers as well as tar. The reciprocity of thermochemical production of tar and release of monomers indicates that the emergence of life would have been a non-monotonic step-by-step process, some or most instantiations of which may have stalled or failed. (In other words, the `winnowing' effect becomes \textit{too} efficient --- in idiom, the baby is thrown out with the bathwater! But we should keep in mind that a pre-biotic world may have been perfectly satisfied with itself; it was not striving to ultimately produce \textit{us}.) Two fictitious but illustrative trajectories suggested by these results are sketched in figure \ref{figure5}. The notional dependent variable `Aliveness' is a term coined by Sutherland \cite{Sutherland:2017} to express the concept that the transition from inanimate matter to life was  a continuous increase in `Aliveness' over time, rather than a discontinuous jump. 
 \begin{figure}[t]
 \centerline{
 \includegraphics[scale=1]{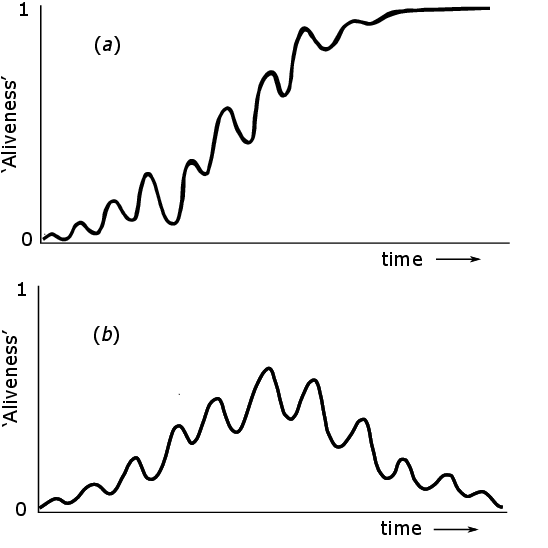}
 }
 \caption{\label{figure5} On a scale of `Aliveness' \cite{Sutherland:2017} where 0 is non-living and 1 is achievement of  proto-cellular life, trajectories are non-monotonic. The troughs represent maximum tarring and the peaks represent maximum monomer desorption. In (a) the troughs and peaks become shallower as enzyme activity evolves, in (b) the process begins well but it succumbs to tarring and collapses.}
 \end{figure}

Stochasticity is central to our results and, we believe, to the emergence of life \cite{Ball:2020,Ball:2021}. Although environmental stochasticity has been investigated in pre-biotic chemical reaction networks \cite{Peng:2020}, it has been in reference to isothermal systems only and has not been modelled as thermal fluctuations, or equivalently, density or chemical potential fluctuations. In our work the normally distributed input thermal fluctuations (equation \eqref{e4}) produce non-normally distributed output fluctuations that enhance monomer desorption and suppress tar formation (table \ref{table0}) and figure \ref{figure4} (a)). The effect is small but significant. In the pre-biotic world large effects are more likely to become catastrophic than constructive --- in this case, too-frequent, violent high-temperature thermal fluctuations might tear apart the monomers faster than they desorb or form, or damage mineral surface sites. Given that the processes are reciprocally linked, we do not want to suppress tar formation \textit{too} much. 

The effects of high temperature stochastic fluctuations may manifest as side-effects downstream of the system defined by equations \eqref{e3}--\eqref{e4}, at neighbouring mineral sites where thermochemically released monomers may be scavenged for recycling. As a hypothetical example, we invoke Mungi's and Rajamani's \cite{Mungi:2015} experimental result that, under dehydration-rehydration cycling in solution, the optimum yield of oligomer from 5$^\prime$AMP occurred at temperatures of 80--90\,C. In a realistic pre-biotic scenario, catalytic mineral surfaces would be integral to the system. Where sites are damaged by high temperature fluctuations, the yield of oligomer against breakdown products and monomer may well be significantly lower.   

On the other hand, the result shown in figure \ref{figure4} (a) supports the findings of previous works \cite{Ball:2020,Ball:2021}: thermal oscillations subject to a stochastic thermal input, equation \eqref{e4}, give rise to a bimodal fluctuation distribution, the higher temperature peak of which is right-weighted. This is described in section 3 as an `inbuilt bias', which effectively means that, over time, synthetic reactions with higher activation energies are favoured downstream, allowing for greater molecular complexity. By contrast, the non-equilibrium \textit{steady-state} fluctuation distribution in figure \ref{figure4} (b) is left-weighted, favouring low activation energy degradation reactions.

\newpage

Evidently pre-biotica were already balancing and responding to constructive and \\degradative processes to optimize growth, evolution and persistence, as biology does to this day. 

Could the presence of tars in certain environments be interpreted as a sign of ancient pre-biotic activity? Given that fully-fledged life evades tar production completely, we could take a broader approach  to labelling selected phenomena  as biosignatures. Even though this term is often misconstrued and still fraught with a lack of precise accepted definition due to probabilistic issues \cite{Barge:2022,Malaterre:2023}, we could develop biosignature criteria applicable to tarry relics. We may interpret a relative \textit{scarcity} of tars as a biosignature of fully-fledged life. On the other hand, the \textit{presence} of tars in specific  settings could be  a probabilistic biosignature of incipient life or of active or extinct prebiotic processes.  If current and future Mars rovers return high-resolution images of subsurfaces we may instruct AI to search the data for patterns of streaky tars. We might expect to learn to distinguish morphological tar patterns due to complex non-equilibrium pre-biotic chemistry from patterns due, perhaps, to chondritic or other relics. 


Our interest in Mars is that is that it may have been a crucible for pre-biotic mineral-organic processes that stalled 3--4 billion years ago before they could progress to fully-fledged cellular life. Certain subsurface environments of Mars  effectively may be fossils of countless `stopped flow' experiments.  

We remarked in section 2 that thermochemically reciprocating oscillations have not been investigated experimentally in a pre-biotic setting. Our modelling results are well-founded on previous published work and provide both motivation and recipe for  experiments. To begin with, we would suggest proof-of-principle experiments in which a slurry of selected finely ground minerals (following W{\"a}chtersh\"auser \cite{Wachterhauser:2006}) and a stream of formose reaction polymeric products are fed through a thermally jacketed micro-CSTR apparatus where the internal temperature is monitored continuously, to observe steady or fluctuating thermal oscillations like those in figure \ref{figure3}. 
Such apparatus is available commercially and is well-known in most chemical engineering, and some physical chemistry, laboratories. Preliminary experiments to obtain  relevant thermokinetic and thermochemical data  would be helpful.  

\section{Conclusions} 
\begin{enumerate}
\item Our dynamical model for a reciprocating thermochemical  subsystem, equations \eqref{e3}--\eqref{e4} is simple, small, and limited as specified, but not too simple, as it yields the useful results shown in figures \ref{figure3} and \ref{figure4} and table \ref{table0} and the insights discussed. Large, complex models are not currently useful for emergence-of-life studies (as they are for, say, weather forecasting) since they are unable as yet to match with historical reality and may inspire a sort of false confidence, leading to anthropocentric self-deception. 

We emphasize that the CSTR paradigm, as we have used in this work, although experimentally well-known and accessible, \textit{is} only a model. An analogy would be the `mouse model' used in biomedical laboratories for studying human diseases. In both cases, since it is not feasible to carry out experiments on, or in, the objects of interest (humans or the pre-biotic world), we use a well-characterized model system for the experiments and interrogate the results to see what they may tell us about the objective system, or what might be the next steps that we could take. More `realistic' systems may be investigated in future to build on these results.

\item We have simulated the  physical messiness --- or thermal variability, as distinguished from isothermal chemical messiness by  \cite{Deamer:2022} --- of an emergence-of-life scenario  by using the stochastic thermal input of equation \eqref{e4} in our model. The resulting output `messiness' --- the right-weighted fluctuation distribution in figure \ref{figure3} (a) --- suggests that time-changing irregularity is not an unwelcome evil but an essential condition for pre-biotic structure to emerge and persist.

\item For amorphous polymeric carbohydrates adsorbed onto a mineral surface, the reciprocating action of our system (figure \ref{figure2} and results) ensures the release of a monomer at the expense of some tar formation. The released monomeric species may, if it is ribose, derivatize  a nucleoside, or be re-cycled into fitter polymers (figure \ref{figure1}). 

\item From these results, we suggest that thermochemical reciprocating action on mineral surfaces provides a quasi-continuous, dynamical backdrop, which may couple to the play of more dramatic pre-biotic chemistry.   
We can surmise that one of the effects, if not purposes, of the emergence of life (towards but not necessarily accomplishing fully-fledged biology) is the avoidance of tarring. 

\end{enumerate}

 \section*{Appendix} 
 \appendix
 
 \begin{table}[h]\caption{\label{table1}Dimensionless variables and parameters  in equations \eqref{e3}--\eqref{e4} defined in terms of physical quantities, which may be expressed in any set of dimensionally consistent units. Subscripts 1 and 2 refer to the alternating monomer detachment/desorption reactions and tar/IOM formation processes respectively, as indicated in figure \ref{figure2}. }

 \centerline{\small
 \begin{tabular}{ll} 
 \hline
 $\alpha$ & $\Delta H_2/\Delta H_1$\\
 $\bar{C}$ & $C_\text{av}E_1/ \left(c_\text{ref}R\left(-\Delta H_1\right)\right)$\\
 $\gamma$ & $G/V\left(z_1c_\text{ref}\right)$\\
 $\varphi$ & $F/\left(Vz_1\right)$\\
 $\ell$ & $LE_1/\left(Vz_1c_\text{ref}R\left(-\Delta H_1\right)\right)$\\
 $\mu$ & $E_2/E_1 $ \\
 $\nu$ & $ z_2/\left(z_1c_\text{ref}\right)$\\
 $\tau$ & $tz_1$\\
 $u$ & $RT/E_1 $\\
$w$ & $c_\text{w}/c_\text{ref}$\\
$x$ & $c_\text{x}/c_\text{ref}$\\
 \hline
 $c_\text{w}$, $c_\text{x}$ & volumetric amounts of dehydration water, desorbed monomer\\
 $ c_\text{ref}$ & a reference concentration\\
 $ C_\text{av} $ & average volumetric specific heat\\
 $ F$ &volumetric flow rate \\
  $E_1$, $E_2$ & activation energies \\
 $G$ & constant rate of substrate supply\\
 $\Delta H_1$, $\Delta H_2$ & reaction enthalpies\\
 $L$ & wall thermal conductance \\
 $R$ & gas constant\\
 $t$ & time\\
 $T$, $T_\text{a}$ & system temperature, constant ambient temperature\\
 $V$ & volume of pore or interfacial zone\\
 $z_1$, $z_2$ & reaction frequencies \\
 \hline
 \end{tabular}}
 \end{table}

\begin{table}[h]\caption{\label{table2} Some germane values for thermokinetic parameters from the literature.  Such data are rare since reaction rates of interest mostly are measured at a single temperature under quasi-equilibrium conditions in solution. Parameter values that were tested in our model are taken from an eclectic range of contexts, mostly not related to pre-biotica in particular. However, even approximate values for reactants with relevant functional groups are useful for our purposes.  PPM is pyrite-pyrrhotite-magnetite, ImpC is the 5$^\prime$-monophosphorimidazolide moiety of cytidine.}
 
 \centerline{\footnotesize
 \begin{tabular}{lccc}
 \hline
 Reaction & $z$ &$E$ (kJ/mol)& Ref.\\
 \hline
 hydrolysis of  ImpC on clay, 60--90 C &$4.58 \times 10^5$ & 61.7 & \cite{Kawamura:2008}\\
 degradation  of oligo- cytosine on clay, 60--90$^\circ$C & $7.4 \times 10^8$ & 93.3 &\cite{Kawamura:2008}\\
 thermal decomposition of threonine in aqueous solution, 113--200$^\circ$C &$1.8\times 10^{12}$ &141&   
 \cite{Vallentyne:1964}\\
 desorption of Acid Blue 193 from Na-bentonite, 20--50$^\circ$C & & 37.5 & \cite{Ozcan:2004}\\
 thermal decomposition of norvaline on  PPM, 150--190$^\circ$C& $ 7.9 \times 10 ^{13}$ & 178 & \cite{McCollom:2013}\\
 hydrothermal decomposition of alanine, 260--320$^\circ$C& $9.4 \times 10 ^{13}$ & 177 & \cite{Zheng:2023} \\
 hydrothermal degradation of alanine, 300--400$^\circ$C& $3.6 \times 10 ^{11}$ & 160 & \cite{Klinger:2007} \\
 hydrothermal degradation of glycine, 300--400$^\circ$C& $1.4 \times 10 ^{12}$ & 156 & \cite{Klinger:2007} \\
 acid hydrolysis of adenosine, 43--58$^\circ$C& $1.7 \times 10 ^{12}$ & 89&\cite{Hevesi:1972}\\ 
 acid hydrolysis of guanosine, 43--75$^\circ$C& $1.3 \times 10 ^{12}$ & 87&\cite{Hevesi:1972}\\ 
 \hline
 \end{tabular}}
 \end{table}
 
\clearpage
\subsubsection*{Author Contributions} Rowena Ball carried out the computational modelling and data analysis and interpretation, graphics, initiated the conception and design of the research, and was the major contributor to drafting and revising the article. John Brindley contributed to the conception and design of the research, data interpretation, and drafting and revising the article. 

 \subsubsection*{Acknowledgements}

\noindent  We thank an anonymous Referee for constructive and encouraging comments, suggestions, and pertinent extra references, which have helped us to clarify the results and discussions, especially for readers from diverse scientific backgrounds.   

\noindent In general, we would like to acknowledge the value of multidisciplinary inputs.  This work owes its conception, input choices, results and interpretations to several disciplines, and is a small demonstration of how multidisciplinary confluences can advance emergence-of-life science. 

\noindent As applied mathematicians, having had long experience working with scientists from other disciplines, including chemical engineering, physics, and physical chemistry, we know that physical insights are gained through recognition and classification of  patterns and periodicities in nature, and we humbly contribute the necessary knowledge of dynamical systems and stability theory, and of statistics. To the chemical engineers, who studied thermochemical oscillations in non-equilibrium flow systems on ever more-highly optimised surfaces and \textit{still} found oscillations, we owe our insights  into the dynamics of processes on organic-geochemical surfaces. Our own, admittedly limited, knowledge of planetary mineral systems we owe to the Earth and planetary scientists.  
Finally, the ingenious experimental work of physical and organic chemists has given us ways by which nature may have synthesized the molecular building blocks of the prebiotic world. 

\subsubsection*{Data Accessibility} Code and data are published at Dryad \href{https://doi.org/10.5061/dryad.kh18932hr}{https://doi.org/10.5061/dryad.kh18932hr}.

\subsubsection*{Ethics} This work did not require ethical approval from a human subject or animal welfare committee.}

\subsubsection*{Funding Statement}

This work was partially funded by Australian Research Council Future Fellowship\\ FT0991007.

\clearpage

\raggedright

\end{document}